# Radiative energy bandgap of nanostructures coupled with quantum emitters around the epsilon-near-zero (ENZ) frequency

Tao Gong,[1,2] Inigo Liberal,[3] Miguel Camacho,[4] Benjamin Spreng,[1] Nader Engheta,[5,*] and Jeremy N. Munday[1,*]

[1]*Department of Electrical and Computer Engineering, University of California, Davis, CA, USA*
[2]*Department of Materials Science and Engineering, University of California, Davis, CA, USA*
[3]*Institute of Smart Cities, Public University of Navarre, Pamplona, Spain*
[4]*Department of Electronics and Electromagnetism, University of Seville, Seville, Spain*
[5]*Department of Electrical and Systems Engineering, University of Pennsylvania, Philadelphia, PA, USA*

Epsilon-near-zero (ENZ) materials have been demonstrated to exhibit unique electromagnetic properties. Here we propose the concept of radiative energy bandgap for an ENZ nanoparticle coupled with a quantum emitter (QE). The radiative emission of the coupled QE-nanoparticle can be significantly suppressed around the ENZ frequency and substantially enhanced otherwise, yielding an effective energy bandgap for radiation. This suppression is effectively invariant with respect to the particle size and is therefore an intrinsic property of the ENZ material. Our concept also heralds an alternative pathway to quench the emission from a QE, which may find potential application in quantum information storage.

The optical properties of a homogeneous non-magnetic material are fundamentally determined by its frequency-dependent permittivity ($\varepsilon = \varepsilon_r + i\varepsilon_i$). For most materials in a broad frequency range $\varepsilon_r$ is either positive or negative, yielding a dielectric or metal-like property, respectively. Nonetheless, there can be narrow frequency bands where $\varepsilon_r$ approaches near-zero values (referred to as epsilon-near-zero, or ENZ), as they are inherent in the material's dispersion. For metals, the ENZ frequencies usually reside in the UV region [1-4]. Other photonic materials, such as transition-metal nitrides and doped semiconducting oxides, have ENZ frequencies in the visible and near-infrared (NIR) regimes, enabling more amenable experimental characterization and potential applications [5-9]. Besides, structured materials (metamaterials, waveguides, or composites), exhibiting effective ENZ-like behaviors, can enable tuning of the ENZ frequency due to resonances or structural mode dispersion through careful design of their geometry and size, however, at the cost of sophisticated fabrication processes and spatial dispersion [10-12].

At the ENZ frequency, the electromagnetic field can be described spatially in a quasi-static limit (while temporally dynamic) with negligible phase advances and strong spatial coherence, resulting in a wealth of fascinating phenomena such as electromagnetic field supercoupling [13,14], resonance pinning [15], perfect optical absorption [16], and ultrafast optical switching [17,18]. In particular, profound modulation of the spontaneous emission from a quantum emitter (QE) has been investigated when coupled with ENZ structures. For instance, the emission is dramatically quenched when a QE is placed inside of an enclosed cavity bound by ENZ walls [19,20]. Contrastingly, strong emission enhancement occurs if the QE is positioned in a nano-channel or waveguide at the cut-off frequency where the ENZ condition holds [21-23], also leading to nonperturbative decay dynamics [24]. However, these emission modulations require the emitter to be placed inside of a closed or semi-closed ENZ structure, bringing about inevitable experimental challenges and limitations.

In this Letter, we theoretically investigate the radiation properties of a perpendicularly oriented QE coupled with an ENZ nanoparticle with realistic material dispersion. We propose and demonstrate numerically the concept of a radiative energy bandgap for a coupled QE-ENZ nanoparticle system. Specifically, the system always exhibits suppressed radiation around the ENZ frequency of the nanoparticle, forming a radiation "stopband", while strong resonance-induced emission enhancement occurs outside the band. Additionally, we find that the bandgap is almost invariant with respect to the structural size. As such, the radiation suppression is mostly determined by the optical properties of the ENZ material (i.e., the ENZ frequency and the dispersion around it) rather than the shape or size of the nanoparticle, in contrast to that induced by photonic or plasmonic resonances. Our results also suggest an alternative route to quench the spontaneous emission from a QE for potential application in quantum information storage.

For a nanoparticle with a given geometry, size, and material, the allowed electromagnetic modes are defined with resonance frequencies and corresponding field distributions. The resonant modes at frequencies with positive $\varepsilon$ correspond to dielectric resonances originating from the displacement currents due to the oscillations of bound electrons, while those at frequencies with negative $\varepsilon$ relate to plasmonic resonances resulting from the oscillation of free electrons at the surface of the particle [25]. Considering the permittivity of a natural material with the conventional Drude-like dispersion, plasmonic resonances are usually located at lower frequencies than the dielectric resonances. Yet, around the ENZ frequency, no resonant modes would exist as the wavelength approaches infinity with quasi-static field spatial distribution across the particle. Consequently, a nanoparticle inherently features discrete mode energy states separated by a primary gap centered around the ENZ frequency, as shown by the schematic in Fig. 1(a).

A direct manifestation of the resonant modes of a nanoparticle is Mie scattering [26]. Figure 1(b) shows the scattering intensity of an ITO nanosphere in the far zone illuminated by a broadband plane wave. The ITO material considered here features an ENZ frequency of ~1.06 eV, corresponding to a free-space wavelength (at ENZ frequency) of ~1.172 μm. We measured its permittivity using spectroscopic ellipsometry over a broad wavelength range (Fig. S1). This material is chosen for analysis because: (1) The material's dispersion nicely spans from positive to negative values in the UV-NIR wavelength regime; (2) the nontrivial loss factor of the material makes the analysis more practically insightful; (3) the material is commercially available, ensuring consistent and repeatable material properties. The intensity distribution of the scattered field reveals that the wavelength associated with the primary dielectric resonance scales approximately with the sphere diameter $D$ as $\lambda \sim n(\lambda)D$, where $n(\lambda) = \sqrt{\varepsilon(\lambda)}$ denotes the wavelength-dependent refractive index of the material [27]. We note that the plasmonic resonance converges to the wavelength (~1.4 μm) that renders $\varepsilon \sim -2$ in the small sphere size limit. This wavelength corresponds to the localized surface plasmon resonance (LSPR). As the size increases, resonance redshifts due to retardation effects [28,29]. Effectively, a wavelength (energy) gap of the scattered photons arises around the ENZ wavelength for a given nanosphere.

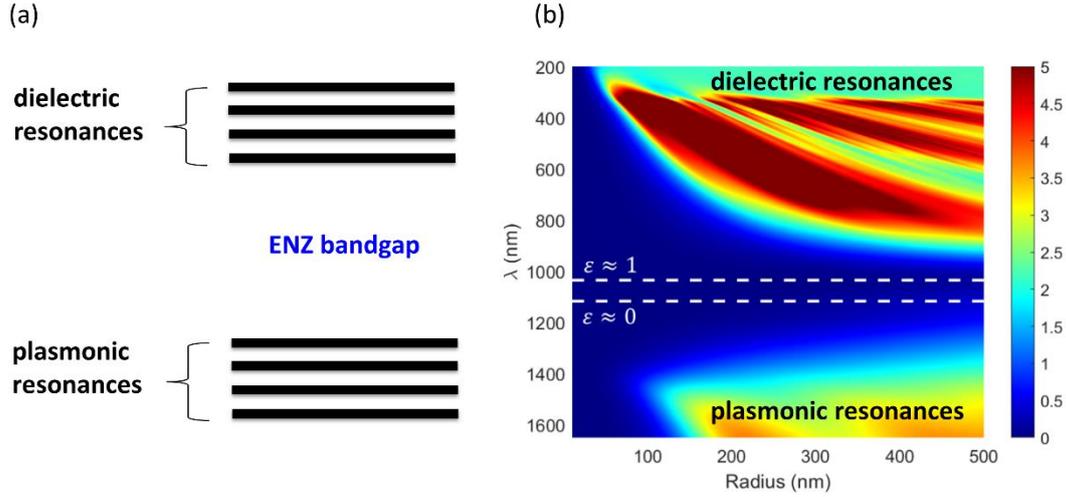

**FIG. 1**. (a) Schematic of the nanoparticle resonant mode energy states and the effective energy bandgap. Around the epsilon-near-zero (ENZ) frequency, no resonant modes can exist, thus an energy "bandgap" arises. (b) Calculated scattering intensity from a nanosphere made of ITO upon the illumination from a broadband plane wave. Strong scattering occurs when plasmonic or dielectric resonant modes are excited, yet scattering is much weaker around the ENZ frequency. A scattered photon wavelength (energy) bandgap therefore is induced.

An alternative manifestation of the bandgap effect is through the coupling with a QE (e.g. an excited fluorescent molecule or a quantum dot). When a QE is placed in proximity to a nanoparticle, a coupled system is created: the presence of the nanoparticle modifies the local density of electromagnetic states at the position of the QE, altering its spontaneous emission rate. In parallel, the presence of the QE may excite resonant modes of the nanoparticle and the nanoparticle will re-radiate (scatter) the photons which may interfere with the QE emission as well. Therefore, as a system, the radiative emission is significantly modified in terms of the spectral profile when compared to an otherwise standalone QE [30-33]. Note that the alteration of the radiative emission depends strongly on the orientation of the QE, the size and shape of the nanoparticle, and the permittivity of the material composing the particle in a nontrivial manner [34-36]. In the past, research on the emission properties of such coupled systems have prevalently been centered on using plasmonic nanoparticles/nanoantennae to enhance or quench the fluorescence and tailor the radiation pattern of molecules whose emission wavelengths are close to the corresponding plasmon resonances of the nanoparticles or nanoantennae [30,31,35,37-41]. Our study here instead focuses on the radiative emission properties of the coupled system in a broader range crossing the ENZ frequency.

Let us examine in brevity how the spontaneous emission from the QE is modified when coupled with a nanoparticle. The inset of Fig. 2(a) shows a perpendicularly oriented QE (modeled as an oscillating dipole) placed at a fixed distance (20 nm) above an ITO nano-platelet with a thickness of 50 nm and varying lateral size. The spontaneous emission from the QE when coupled to the ITO platelet is substantially enhanced (by a factor of ~1300), with the maximum enhancement occurring at a wavelength of ~1.29 μm. This maximum translates to the resonance frequency of the propagating surface plasmon polariton (SPP) where $\varepsilon \sim -1$, particularly when the lateral size of the plate is sufficiently large (Fig. 2(a)). This enhancement can be attributed to the resonant energy transfer from the QE to the SPP mode of the nanoplatelet [37]. We also note that the enhancement saturates when the lateral size goes beyond 200 nm. This behavior is due to the non-negligible loss in the material which limits the propagation of the SPP within ~200 nm laterally; a further increase in the lateral size does not increase the coupling strength from the QE to the polariton.

The radiative emission from the coupled QE-nanoplatelet system undergoes either an enhancement or suppression, depending on the wavelength of interest. Intriguingly, the radiation is notably suppressed around the ENZ frequency irrespective of the lateral size of the platelet, although the optimal suppression and the corresponding wavelength varies slightly (Fig. 2(b)). The suppression of radiation can be understood as a result of the destructive interference of the QE and the induced dipole in the platelet. In an ideal scenario, the boundary condition at the interface between a bulk ENZ material and vacuum is equivalent to that between vacuum and a perfect magnetic conductor (PMC), where an image dipole is induced inside the material with a 180 degree out-of-phase oscillation along the perpendicular direction. However, the non-monotonic variation of the suppression as well as its corresponding wavelength is indicative of the deviation from the perfect image charge model because of the convoluted interplay between the material's loss, finite thickness and finite lateral size of the platelet, and the finite distance between the QE and the platelet. As a comparison, an ideal ENZ platelet with negligible loss features a smooth convergence of the wavelength of optimal suppression to the ENZ wavelength (Fig. S2(a)). Figure 2(c) shows that although the radiation from the coupled system is suppressed at the ENZ frequency, emission from the QE is increased by more than 200 times. This behavior indicates that most of the energy transferred from the QE is dissipated nonradiatively at the ENZ frequency due to considerable loss factor inside the material even if no resonant mode is excited. The inset shows the electric field distribution in the platelet with a lateral size of 100 nm at two different wavelengths. At the SPP resonance wavelength (~1.3 μm), the field is mostly concentrated at the top ITO-air interface because the SPP propagates near the surface; however, the field is distributed throughout the slab at the ENZ wavelength (~1.172 μm). We note here at the ENZ frequency the attenuation of the field in the vertical direction is attributed to the loss of the material, not due to surface wave effect as for the SPP resonance. For comparison, the field distribution in an ideal ENZ platelet is relatively uniform vertically across the slab, as shown in Fig. S2(a).

The full spectra of the radiated power from the coupled QE-platelet with varying lateral size is shown in Fig. 2(d). We observe a narrowband of radiation suppression located around the ENZ frequency, with a maximum of nearly 50% suppression. This behavior signifies a radiation energy "bandgap" where much fewer photons can be radiated out. In fact, if the ITO material is replaced with an ideal ENZ material, complete suppression is attainable (Fig. S2(b)). We further note that the maximum radiation enhancement takes place around the SPP resonance (~1.3 μm), but the corresponding wavelength progressively blueshifts as the size increases. This behavior indicates that the maximum energy transfer from the QE to the nanoplate at the SPP resonance does not necessarily convert into optimal radiation, as the radiation is due to the outcoupling of the surface plasmon mode into radiation. At the SPP resonance the outcoupling is not the most efficient since most of the energy is dissipated nonradiatively. For shorter wavelengths ($\varepsilon > 0$), the radiation enhancement is attributed to the primary dielectric resonant mode, and higher-order resonances cannot be excited with such a thin slab. In small size limit, the enhancement is consistently small across all wavelengths, in agreement with the scaling law of the radiative rate with the nanoparticle size.

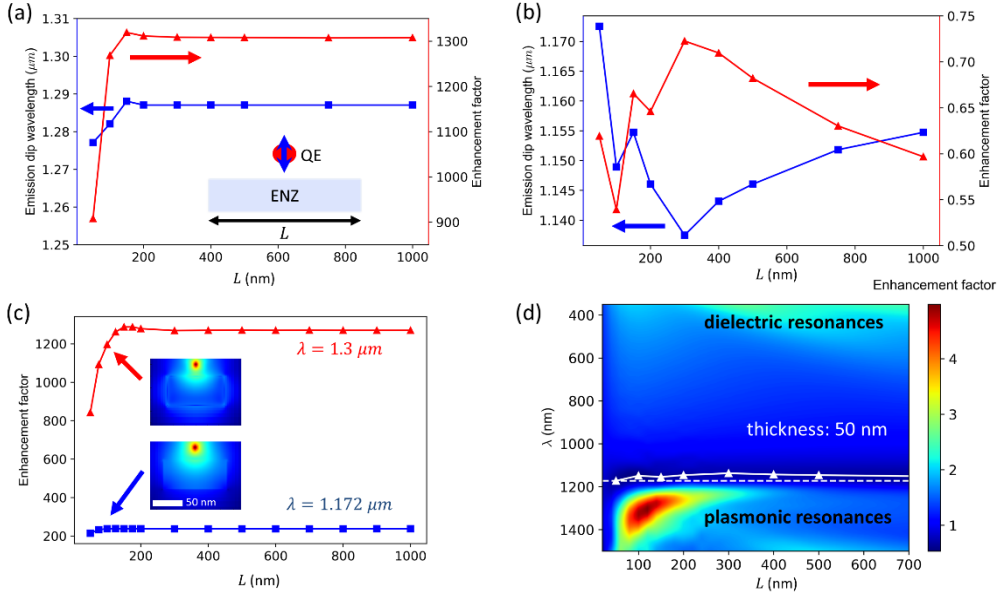

**FIG. 2.** Radiative emission enhancement of a perpendicularly oriented QE coupled with an ITO nanoplatelet of varying lateral size. The QE is placed 20 nm above the 50 nm-thick plate. (a) Optimal enhancement of the spontaneous emission from the QE (red, right axis) and its corresponding wavelength (blue, left axis) as a function of the platelet lateral size $L$. Inset: schematic of the structure. (b) Optimal suppression (minimum enhancement) of the radiation from the coupled QE-platelet (red, right axis) and its corresponding wavelength (blue, left axis) as a function of $L$. (c) Enhancement factor of the spontaneous emission from the QE as a function of $L$ at two specific wavelengths: 1.172 μm (ENZ wavelength, blue curve) and 1.3 μm (SPP resonance, red curve). Inset: the electric field intensity distribution inside the platelet. (d) Enhancement spectra of the radiative emission from the coupled QE-platelet for varying $L$. A marked radiation suppression is observed within a wavelength band around the ENZ wavelength (the white dashed line). The white solid line denotes the wavelengths corresponding to the optimal radiation suppression for different $L$.

The outcoupling of the excited mode into radiation also hinges on the plate thickness. Here we fix the lateral size of the plate to be 100 nm (Fig. 3(a,c)) and 500 nm (Fig. 3(b,d)), respectively, and vary the plate thickness. In a similar vein, we observe a radiation suppression band near the ENZ frequency, with an optimal suppression above 50% (maximum at 80%). Moreover, the wavelength at which optimal suppression occurs approaches the ENZ frequency as the plate thickens (Fig. 3(c) and 3(d)). This effect can be understood as the better accuracy of the image dipole model for thicker ENZ plate as discussed above. In addition, the bandgap is almost invariant with respect to the thickness of the platelet (Fig. 3(a,b)). At short wavelengths, the dielectric resonances enhance the radiation more notably with larger and thicker plates, and more resonant modes of higher orders are induced (Fig. 3(b)). For the maximum radiation enhancement around the plasmon resonance frequency, we note that in both cases the wavelength of optimal enhancement converges to ~1.32 μm, slightly redshifted from the SPP resonance (Fig. S3). This behavior indicates that the most efficient outcoupling of the plasmon mode into radiation is relatively agnostic to the thickness, especially when the lateral size is large. This effect can be understood as the induced dipole and the propagating field being confined close to the top surface. For the plate with a smaller lateral size, larger field can be found down the side walls, rendering the radiation more dependent on the thickness than from the one with a larger lateral size.

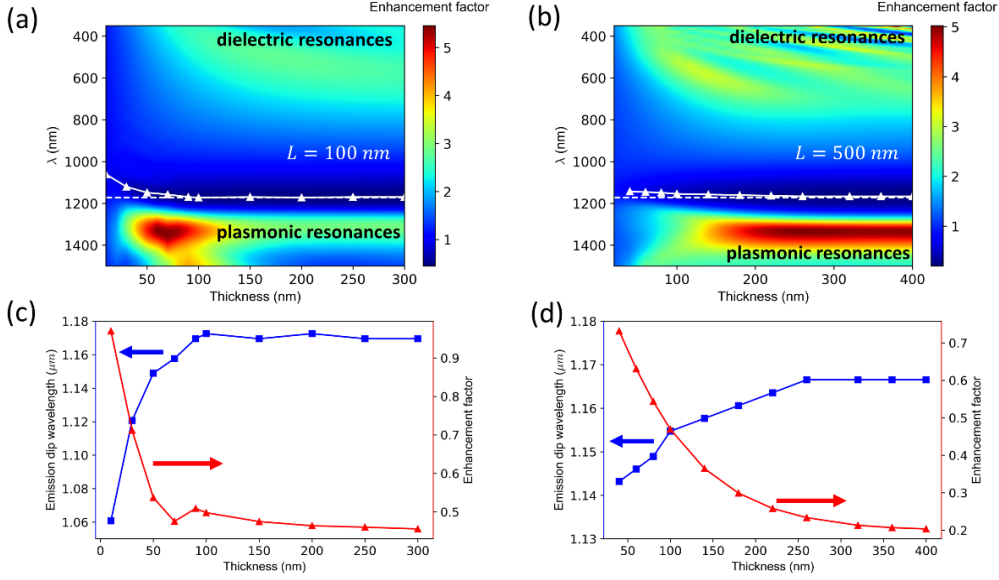

**FIG. 3.** Radiative emission enhancement of a perpendicularly oriented QE coupled with an ITO nanoplatelet of varying thickness. The QE is placed 20 nm above the plate. The lateral size of the plate is (a,c) 100 nm and (b,d) 500 nm. (a) and (b): Enhancement spectra of the radiative emission from the coupled QE-platelet for varying platelet thickness. A marked radiation suppression is observed within a wavelength band around the ENZ frequency (the white dashed line) for both lateral sizes, with the maximum suppression for greater than 80%. The white solid line denotes the wavelengths corresponding to the optimal radiation suppression for different thicknesses. (c) and (d): Optimal suppression (minimum enhancement) of the radiation from the coupled QE-platelet and its corresponding wavelength as a function of the platelet thickness. The wavelength of optimal suppression converges to the ENZ frequency as the plate is thickened.

Let us simplify the geometry to a sphere, with only one parameter to be considered: sphere radius. We first examine the emission modulation of the QE. Figure 4(a) shows that the enhancement from the QE can reach ~1600 when coupled to the plasmon resonance of the ITO sphere. The enhancement peak occurs at a wavelength that shifts gradually from the LSPR at ~1.41 μm to the SPP resonance at ~1.3 μm as sphere radius increases. This effect can also be visualized in Fig. 4(b) where a strong emission enhancement band is observed in the neighborhood of the plasmonic resonances. Concurrently, the radiation from the coupled QE-nanosphere is also modified when compared to an isolated emitter. Figure 4(c) shows the non-monotonic change of the corresponding wavelength for the maximum radiation suppression. As expected, the larger the sphere, the closer the wavelength is to the ENZ frequency for the optimal suppression. Remarkably, the suppression can reach nearly 90% when sphere radius goes up to ~300 nm. Analytically, the radiation suppression from the coupled QE-nanosphere around the ENZ frequency can be approximated. To first-order, the radiation enhancement factor in the near-field coupling regime follows [42]:

$$\frac{\gamma_{rad}}{\gamma_0} \sim 1 + \frac{4}{R^3} Re(\alpha_1) + \frac{4}{R^6}|\alpha_1|^2$$

where $\gamma_{rad}$ is the radiation rate of the coupled system, $\gamma_0$ is the spontaneous emission rate of a standalone QE, $R$ is the radius of the nanosphere, and $\alpha_1 \sim \frac{\varepsilon_S - 1}{\varepsilon_S + 2} R^3$ is the polarizability of the nanosphere assuming a dipolar response because the dipolar mode represents the highest scattering cross-section. For an ideal ENZ material, $\varepsilon_S \sim 0$, $\alpha_1 \sim -1/2 R^3$, and therefore $\frac{\gamma_{rad}}{\gamma_0} \sim 0$. Further, the edge of the radiation suppression band is plotted to exemplify the invariance of the bandgap with respect to the size of the sphere (Fig. 4(d), two green dashed lines). Note as a conceptual example here we define the radiation suppression band to be such that the enhancement factor is smaller than 1.

Outside the suppression band, we make three observations: Firstly, the radiation map resembles the Mie scattering map in Fig. 1(b). Secondly, compared with Fig. 4(b), the maximum enhancement of the radiation from the coupled QE-sphere due to plasmonic resonances occurs at a wavelength redshifted from the exact LSPR (around ~1.41 μm) due to the dipole-dipole interaction and retardation effects [42]. Thirdly, the dielectric resonances induce much smaller enhancement factors for the spontaneous emission from the individual QE when compared to the plasmonic resonances. This effect is caused by the much weaker coupling of the QE emission into the nonradiative decay in the sphere when it is excited through the dielectric resonant modes. Instead, most of the enhancement of the QE emission is transferred to the enhanced overall radiation of the coupled system. This behavior is the exact opposite to what happens at the plasmonic resonances, where nonradiative dissipation in the nanosphere dominates due to the Ohmic loss. A direct comparison between the enhancement of the spontaneous emission of the QE and of the coupled system can be seen in Fig. S4.

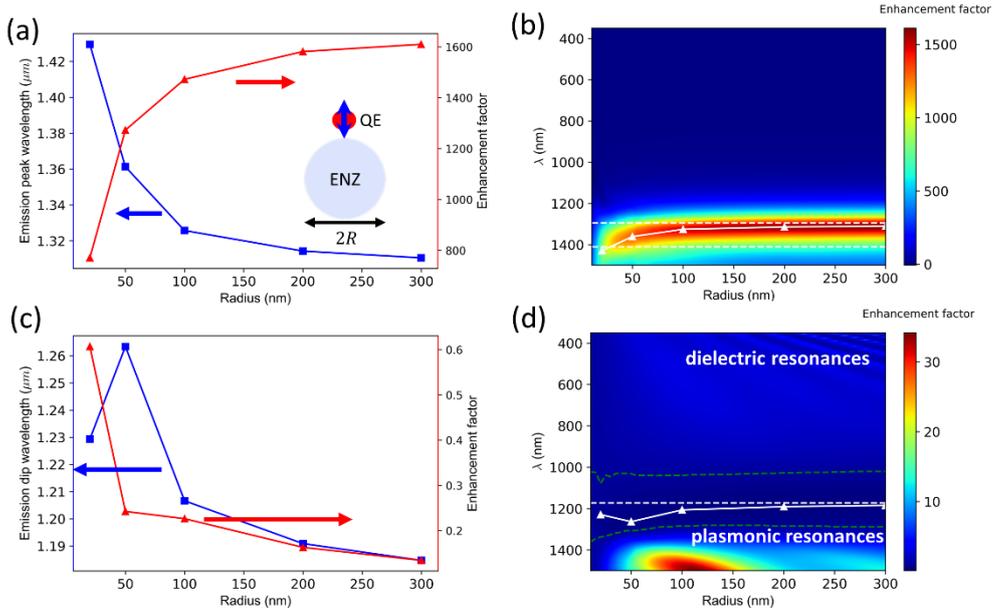

**FIG. 4.** Radiative emission enhancement of a perpendicularly oriented QE coupled with an ITO nanosphere of varying radius. The QE is placed 20 nm above the sphere. (a) Optimal enhancement of the spontaneous emission from the QE and its corresponding wavelength as a function of the sphere radius. (b) Enhancement spectra of the emission from the QE for varying radius of sphere. Optimal coupling from the QE to the sphere is located within a wavelength band in the neighborhood the plasmon resonances, progressively transitioning from LSPR with small sphere size to SPP resonance with large sphere size (the two white dashed lines denote the wavelengths of the two resonances). The white solid line denotes the wavelengths corresponding to the optimal radiation enhancement for different radius. (c) Optimal radiation suppression (minimum enhancement) from the coupled QE-sphere and its corresponding wavelength as a function of the sphere radius. The wavelength for the optimal suppression converges to the ENZ frequency with increasing sphere size. (d) Enhancement spectra of the radiation from the coupled QE-sphere for varying sphere radius. A marked suppression band is observed around the ENZ wavelength (the white dashed line). The white solid line denotes the wavelengths corresponding to the optimal radiation suppression for different radius. The two green dashed lines denote the "band edge" of the radiation suppression band.

In conclusion, we have put forward the concept of a radiative energy bandgap for a coupled QE-ENZ nanoparticle system. A radiation energy gap is always found close to the ENZ frequency, beyond which strong radiation enhancement arise from plasmonic or dielectric resonant mode excitation. Our work also indicates that in order to suppress the radiation, the optical properties of the material are more important than the geometry or size of the nanoparticle. Therefore, our proposed concept suggests an alternative route to quench spontaneous emission, which may open new opportunities in the field of molecular fluorescence modulation and quantum information technology.

The authors wish to acknowledge financial support from the Defense Advanced Research Program Agency (DARPA) QUEST program No. HR00112090084.

# Supplemental Material for: Radiative energy bandgap of nanostructures coupled with quantum emitters around the epsilon-near-zero (ENZ) frequency

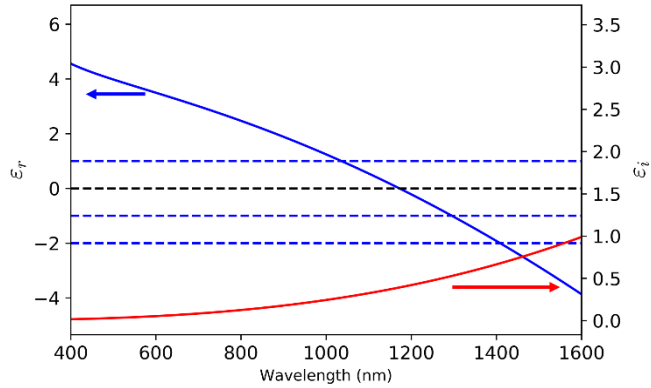

FIG. S1. Spectroscopic Ellipsometry (SE) – characterized dielectric function of the commercially purchased ITO material of interest in this Letter. The real and imaginary part of the dielectric function ($\varepsilon_r$ and $\varepsilon_i$) are represented by the blue and red curve, respectively. The zero-crossing wavelength (ENZ) of $\varepsilon_r$ is around 1.172 μm, as indicated by the black dashed line. The other blue dashed lines represent $\varepsilon_r = 1$ (~1.04 μm), $\varepsilon_r = -1$ (~1.29-1.3 μm) and $\varepsilon_r = -2$ (~1.4-1.41 μm), respectively.

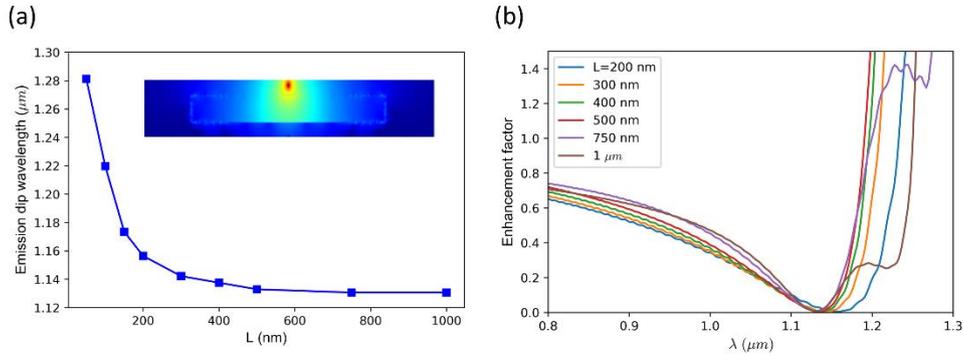

FIG. S2. Radiative emission enhancement of a perpendicularly oriented QE coupled with an ENZ nanoplatelet of varying lateral size. The QE is placed 20 nm above the plate. The plate is 50-nm thick and comprises a low-loss ENZ material described with Drude-model parameters: $\omega_p = 1.1\ eV$, $\gamma_p = 0.01\ eV$, which corresponds to an ENZ frequency with the free-space wavelength of ~1.13 μm. (a) The wavelength that corresponds to the optimal radiation suppression as a function of the platelet lateral size L. It converges to the ENZ frequency. Inset: distribution of the electric field intensity inside the platelet. The lateral size is 500 nm. (b) Emission enhancement spectra of the coupled QE-platelet with different lateral size. Almost complete quenching of radiation can be achieved.

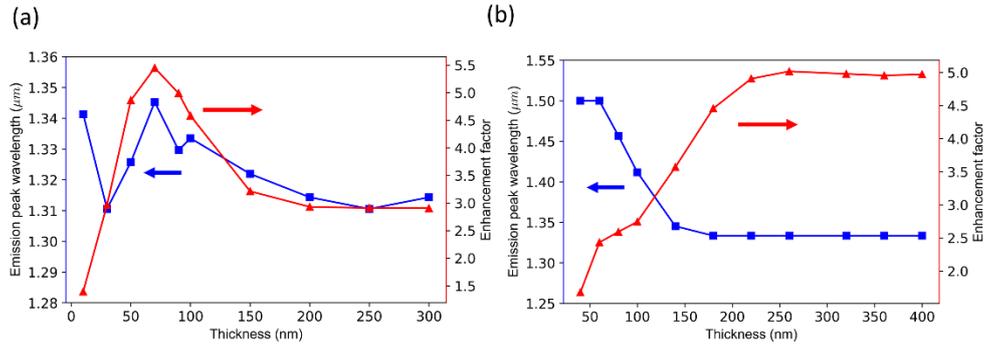

FIG. S3. Variation of the maximum enhancement of the radiation from the coupled QE-platelet and its corresponding wavelength as a function of the platelet thickness for a lateral size of (a) 100 nm and (b) 500 nm. In both cases the wavelength of optimal enhancement all converges to >1.3 μm, slightly redshifted from the SPP resonance.

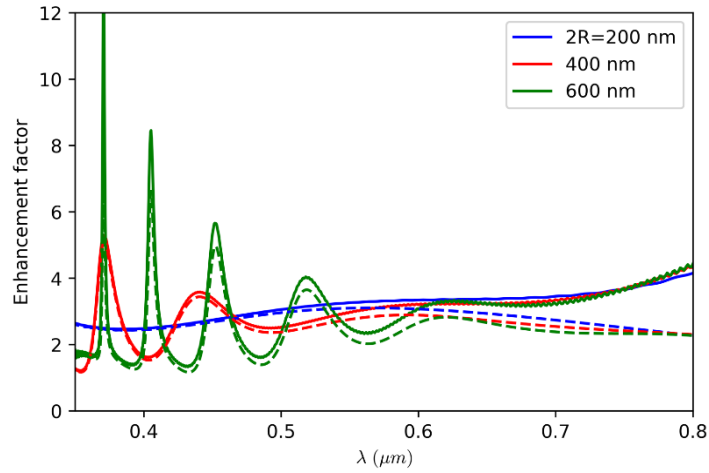

FIG. S4. Enhancement of the spontaneous emission from the QE (solid lines) and the radiation from the coupled QE-ITO sphere hybrid system (dashed lines) in the visible wavelength caused by the dielectric resonances. On resonances, most of the enhancement from the QE is coupled into the overall radiation. The non-radiative dissipation is much less significant.